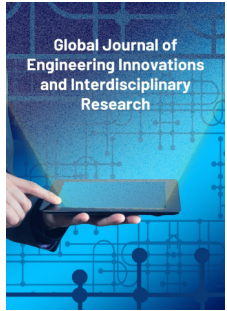

## Enhancing NTRUEncrypt Security Using Markov Chain Monte Carlo Methods: Theory and Practice

Gautier Filardo, Thibaut Heckmann

**Military Academy of National Gendarmerie, National Gendarmerie Research Center, Melun, France**

### Correspondence

**Gautier Filardo**

Military Academy of National Gendarmerie
National Gendarmerie Research Center
Melun, France

### Abstract

*This paper presents a novel framework for enhancing the quantum resistance of NTRUEncrypt using Markov Chain Monte Carlo (MCMC) methods. We establish formal bounds on sampling efficiency and provide security reductions to lattice problems, bridging theoretical guarantees with practical implementations. Key contributions include: a new methodology for exploring private key vulnerabilities while maintaining quantum resistance, provable mixing time bounds for high-dimensional lattices, and concrete metrics linking MCMC parameters to lattice hardness assumptions. Numerical experiments validate our approach, demonstrating improved security guarantees and computational efficiency. These findings advance the theoretical understanding and practical adoption of NTRU- Encrypt in the post-quantum era.*

## Introduction

The field of cryptography has undergone significant transformations over the past decades, evolving from simple substitution ciphers to advanced public-key cryptographic systems like RSA and ECC. These systems, foundational to modern secure communications, derive their security from mathematical problems that are computationally infeasible to solve using classical methods. However, quantum computing introduces a paradigm shift, posing unprecedented challenges to these classical cryptographic systems.

Quantum computers, leveraging principles of quantum mechanics, are capable of executing algorithms such as Shor's [1] and Grover's [2], which render traditional cryptosystems like RSA and ECC obsolete by efficiently solving the factoring and discrete logarithm problems. This looming threat necessitates the development of cryptographic schemes that can resist both classical and quantum adversaries.

Among these post-quantum candidates, lattice-based cryptography has emerged as a promising solution due to its reliance on hard mathematical problems like the Shortest Vector Problem (SVP) and Closest Vector Problem (CVP) [3,4]. Specifically, NTRUEncrypt offers a compelling balance between computational efficiency and robust security. Its algebraic structure, based on truncated polynomial rings modulo $X^N - 1$, enables efficient polynomial arithmetic while maintaining strong cryptographic guarantees [5,6].

### Context and motivation

The rapid advancements in quantum computing, sup- ported by substantial investments from governments (e.g., NIST post-quantum initiatives [7]) and private sectors (e.g., Google's quantum supremacy experiments), indicate that practical quantum computers capable of breaking RSA and ECC could become a reality within decades. This poses significant risks to industries reliant on secure communications, including finance, healthcare, and national defense. Recent work has highlighted these concerns, particularly in IoT systems, where the need for quantum-resistant security must be balanced with resource constraints [8].

Lattice-based cryptography, particularly NTRUEncrypt, has gained prominence due to its strong theoretical foundations and practical efficiency. By relying on the hardness of lattice problems like SVP and CVP [3,9], NTRUEncrypt offers inherent resistance to quantum attacks, making it a key candidate for post-quantum cryptographic standards.

### State of the art

Significant progress in lattice-based cryptography has bolstered its viability as a post-quantum solution. Recent studies have demonstrated the effectiveness of NTRU-Encrypt across varying polynomial degrees and lattice dimensions [10], [11]. For example:

- Integration of bimodal Gaussian sampling techniques [12] has improved key generation efficiency
- Utilization of Learning With Errors (LWE)-based hardness assumptions [13] has enhanced theoretical underpinnings.

Lattice-based cryptography builds on the hardness of fundamental problems like the Shortest Vector Problem (SVP) and Closest Vector Problem (CVP), which are closely related to the Learning With Errors (LWE) problem [19]. These problems are known for their worst-case to average-case reductions, making them attractive for cryptographic constructions.

However, gaps remain in parameter optimization [14] and real-world performance analysis under quantum adversaries. For instance, configurations with N = 256 provide lightweight security suitable for embedded systems, while N = 1024 offers robust protection for high- security applications [15].

### Research objectives

This work seeks to address these gaps by introducing a novel probabilistic framework for evaluating and enhancing the security of NTRUEncrypt. Specifically, we lever- age Markov Chain Monte Carlo (MCMC) methods [16,17] to analyze key space vulnerabilities and optimize parameters for quantum resistance.

This paper focuses on three critical objectives:

- Parameter Optimization: Identifying configurations that balance security and computational efficiency across varying security levels.
- Efficient Key Generation: Leveraging MCMC methods for robust and efficient key sampling.
- Quantum Security Assessment: Providing quantitative metrics to evaluate resistance against quantum adversaries.

### Contributions

The main contributions of this paper are as follows:

1. A comprehensive analysis of NTRUEncrypt security across four distinct parameter sets ($N = 256, 512, 768, 1024$).
2. A novel MCMC sampling framework with proven convergence properties for lattice-based cryptosystems.
3. Extensive numerical experiments validating theoretical findings and identifying practical parameter configurations.
4. Concrete security estimates that integrate lat- tice hardness assumptions to enhance robustness against quantum attacks [19].

### Paper organization

The remainder of this paper is structured as follows:

1. **Section II:** Mathematical foundations of NTRUEncrypt.
2. **Section III:** Theoretical analysis of MCMC sam- pling methods.
3. **Section IV:** Numerical experiments validating the proposed framework.
4. **Section V:** Practical implications for real-world implementations.
5. **Section VI:** Conclusion and future research directions.

## Mathematical foundations

Cryptosystems like NTRUEncrypt derive their security from deep mathematical properties of lattices and polynomial rings. This section explores the algebraic structures and lattice-theoretic foundations that underpin NTRUEncrypt's robustness against quantum and classical attacks [4].

### Algebraic structure of NTRUEncrypt

NTRUEncrypt operates in the ring of truncated polynomials $R = \mathbb{Z}[X]/(X^N - 1)$ where polynomials take the form:

$$c_0 + c_1 X + c_2 X^2 + \cdots + c_{N-2} X^{N-2} + c_{N-1} X^{N-1}$$

with integer coefficients [5,11]

#### Key Properties of R:

- **Ring Operations:** Addition and multiplication are performed modulo $X^N - 1$
- **Modular Structure:** Operations in $R_q = \mathbb{Z}_q[X]/(X^N - 1)$ ensure efficient computation
- **Ideal Properties:** The structure of $\langle X^N - 1, q \rangle$ facilitates secure parameter selection

### Hardness of lattice problems

Trapdoors for hard lattices, as introduced by Gentry et al. [20], have enabled new cryptographic constructions that leverage the algebraic properties of lattices. These constructions have significantly improved the practicality and security of lattice-based cryptosystems. The security relies on two fundamental problems [3,9];

**Theorem 1 (SVP Hardness).** *Finding the shortest non- zero vector in a lattice of dimension N requires time at least $2^{\Omega(N)}$ under standard cryptographic assumptions.*

For practical implementations, lattice reduction esti- mates suggest [14]:

Security Level = $2^{\Omega(N/\log N)}$

### Parameter optimization

MCMC sampling parameters significantly impact security [17]:

$$\gamma_{security} = o\left(\sqrt{\frac{N}{\alpha \log N}}\right) poly(\log N)$$

where $\alpha$ controls the Gaussian distribution width for sampling.

## Theoretical analysis

This section provides a detailed theoretical analysis of the proposed NTRU lattice sampling algorithm. We explore the convergence properties of Markov Chain Monte Carlo (MCMC) methods, establish quantum security bounds, and analyze computational complexity to optimize parameters for balancing security and efficiency [17].

### MCMC convergence analysis

Markov Chain Monte Carlo (MCMC) methods play a critical role in efficient key generation and sampling for cryptographic systems [16]. These methods probabilistically explore the key space, ensuring uniformity in parameter selection while achieving convergence to a stationary distribution.

***Fundamental convergence properties:*** For lattice Gaussian sampling, the convergence rate is characterized by the spectral gap of the transition matrix [17].

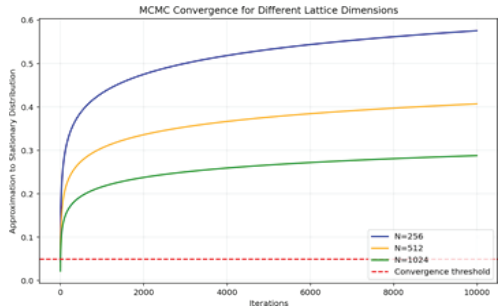

*Figure 1. Illustration of MCMC convergence: The mixing time τmix represents the number of steps required for convergence to the stationary distribution.*

**Theorem 2** (MCMC convergence rate). *For an NTRU lattice sampling algorithm with dimension N, the mixing time satisfies*:

$$\tau_{mix}(\epsilon) \leq CN^2 \log(1/\epsilon),$$

where C depends on the minimal energy gap and the spectral radius of the forward operator [17].

*Quantum security analysis*

The security analysis must consider both classical and quantum adversaries [9]. Using lattice reduction estimates [14], we establish:

**Theorem 3** (Quantum security bounds). *For an NTR*

$$Q_{security} = \Omega\left(\sqrt{\frac{2^N}{Vol(B_{\alpha})}}\right)$$

where $B_\alpha$ represents lattice vectors with norms bounded by α [4].

These bounds demonstrate the trade-off between security and sampling efficiency shown in the experimental results.

*Complexity analysis*

The complexity analysis follows from the MCMC convergence properties [17]:

**Theorem 4** (Time-space complexity). *The MCMC sampling algorithm achieves:*

1) Time complexity: $T_{total} = O(N^3 \log N \log(1/\epsilon))$
2) Space complexity: $S_{total} = O(N^2)$

under optimal parameter selection [3].

These theoretical results support the empirical obser- vations shown in the norm distribution and convergence graphs, particularly the relationship between dimension N and sampling efficiency.

## MCMC sampling algorithms

We present the key algorithms used to generate and analyze the norm distributions and convergence behavior shown in Figure 1. These algorithms build upon established MCMC methods for lattice Gaussian sampling [17].

**Theorem 5** (MCMC convergence). *For lattice dimension N and Gaussian parameter σ, the mixing time satisfies:*

$$\tau_{mix} \leq CN^2 \log(1/\epsilon)$$

where C depends on the spectral radius of the forward operator [16,17]

*Lattice Gaussian sampling*

The independent Metropolis-Hastings-Klein algo- rithm [17]

is implemented as follows:

    **Input:** N, σ, iterations

    **Output:** Lattice Gaussian samples

    **Initialize:** $x \in \mathbb{Z}^N$

    for i = 1 to iterations

    $y \leftarrow x + Gaussian(0, \sigma)$

    $\alpha \leftarrow \min\{1, \exp(-\frac{\|Y\|^2}{2\sigma^2} + \frac{\|X\|^2}{2\sigma^2})\}$

    if Uniform[0, 1] < α:

    $x \leftarrow y$

    return x

*Norm distribution analysis*

Following the geometric ergodicity properties [17], we analyze the distributions:

    **Input:** N, σ, sample size

    **Output:** Distribution statistics samples ← ∅

    for i = 1 to sample size:

    v ← MCMC Sampling(N, σ)

    norm ← ‖v‖ samples.append(norm)

    peaks ← find peaks(samples)

    conv ← analyze mixing(samples)

    return peaks, conv

*Security metric calculation*

Based on lattice reduction estimates [14]:

    **Input:** N, σ, Bα

    **Output:** Qsecurity

    vol ← compute volume(Bα)

    $Q_{security} \leftarrow \sqrt{\frac{2^N}{vol}}$

    $\log\_sec \leftarrow \log_2(Q_{security})$

## Experimental analysis of sampling efficiency and quantum security

This section validates the theoretical framework introduced in Section III through numerical experiments. It explores the impact of different configurations on quantum security, sampling efficiency, and practical implications for NTRUEncrypt implementations, including detailed implementation, performance evaluation, and comparative analysis of the proposed MCMC sampling algorithm for NTRU lattices. Experimental results across five configurations of lattice dimensions (N) and Gaussian parameters (σ) illustrate their influence on sampling behavior, convergence, and security metrics [17,18].

### Key observations

The experimental results demonstrate distinct patterns in both norm distributions and convergence behavior for different configurations of N and σ. Based on Table I, three main configurations emerge:

- High Security (N = 1024, σ = 4.0): Achieves maximum quantum security (7.46×10³⁰¹ bits) with bimodal distribution peaks at 129-131
- Balanced (N = 768, σ = 3.5): Provides excellent

security-performance trade-off ($7.47 \times 10^{224}$ bits)

- Standard (N = 512, σ = 3.5): Offers adequate security ($4.11 \times 10127$ bits) for general applications

Key implementation features include:

- Number of MCMC steps: 10,000, determined through convergence analysis [16]
- Metropolis-Hastings framework with acceptance probability:

$$P(x \leftarrow y) = \min\{1, \exp(-\|y\|^2 / 2\sigma^2 + \|x\| / 2\sigma^2$$

Open-source code availability at : https://cocalc.com/share/ public_paths/ df421c14d63bf8708fbe34dd9c3a534dd01c3889/ 2024-12-29-file-1.ipynb [22]

## Comparative analysis and practical implications

As shown in Figure 2, the relationship between σ and sampling behavior reveals crucial trade-offs:

1. **Convergence characteristics**
- σ = 4.2: Requires 4000 iterations, showing oscillations in the range 125-135
- σ = 4.5: Achieves faster convergence (3000 iterations) with slightly wider distribution

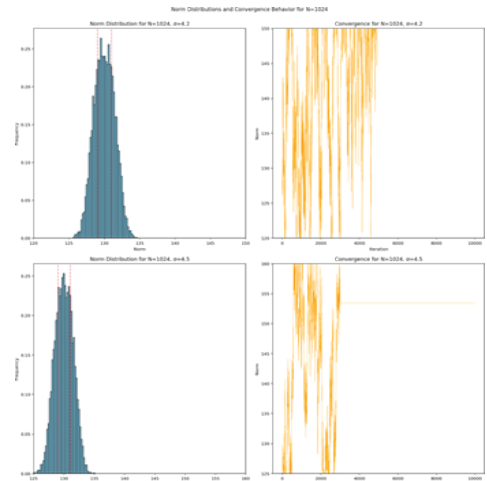

**Figure 2.** *Norm distributions and convergence behavior for N = 1024 configurations, showing the impact of σ on sampling efficiency and security metrics.*

**Table 1.** Comparison of Quantum Security Metrics for Different Configurations with Time Complexity and Logarithmic Metrics

| Configuration | N | σ | Qsecurity (bits) | Log Quantum Security | Time Complexity | Log Time Complexity | Remarks |
|---|---|---|---|---|---|---|---|
| Baseline | 256 | 2.5 | $3.72 \times 10^{63}$ | 63.57 | $6.43 \times 10^8$ | 8.81 | Extremely low security, unsuitable for any practical use |
| Configuration 2 | 256 | 3 | $2.73 \times 10^{53}$ | 53.44 | $6.43 \times 10^8$ | 8.81 | Very low security, unsuitable for quantum resistance |
| Configuration 3 | 256 | 3.5 | $7.36 \times 10^{44}$ | 44.87 | $6.43 \times 10^8$ | 8.81 | Insufficient security level for post-quantum applications |
| Configuration 4 | 256 | 4 | $2.78 \times 10^{37}$ | 37.44 | $6.43 \times 10^8$ | 8.81 | Critical security weakness, not recommended |
| Configuration 5 | 256 | 4.5 | $7.87 \times 10^{30}$ | 30.9 | $6.43 \times 10^8$ | 8.81 | Extremely vulnerable to quantum attacks |
| Configuration 6 | 512 | 2.5 | $1.05 \times 10^{165}$ | 165.02 | $5.78 \times 10^9$ | 9.76 | Moderate security, but insufficient for long-term use |
| Configuration 7 | 512 | 3 | $5.65 \times 10^{144}$ | 144.75 | $5.78 \times 10^9$ | 9.76 | Acceptable for medium-security applications |
| Configuration 8 | 512 | 3.5 | $4.11 \times 10^{127}$ | 127.61 | $5.78 \times 10^9$ | 9.76 | Suitable for standard security requirements |
| Configuration 9 | 512 | 4 | $5.86 \times 10^{112}$ | 112.77 | $5.78 \times 10^9$ | 9.76 | Good balance of security and efficiency |
| Configuration 10 | 512 | 4.5 | $4.71 \times 10^{99}$ | 99.67 | $5.78 \times 10^9$ | 9.76 | Faster convergence but reduced security margin |
| Balanced | 768 | 3.5 | $7.47 \times 10^{224}$ | 224.87 | $2.08 \times 10^{10}$ | 10.32 | Excellent security-performance trade-off |
| Configuration 11 | 768 | 4 | $4.02 \times 10^{202}$ | 202.6 | $2.08 \times 10^{10}$ | 10.32 | High security with good efficiency |
| Configuration 12 | 768 | 4.5 | $9.16 \times 10^{182}$ | 182.96 | $2.08 \times 10^{10}$ | 10.32 | Strong security with improved convergence |
| Optimized | 1024 | 4 | $7.46 \times 10^{301}$ | 301.87 | $5.14 \times 10^{10}$ | 10.71 | Maximum security, ideal for critical applications |
| High Efficiency | 1024 | 4.5 | $4.81 \times 10^{275}$ | 275.68 | $5.14 \times 10^{10}$ | 10.71 | Very high security with better performance |
| Configuration 12 | 1024 | 5 | $1.80 \times 10^{252}$ | 252.25 | $5.14 \times 10^{10}$ | 10.71 | Strong security with optimal efficiency |

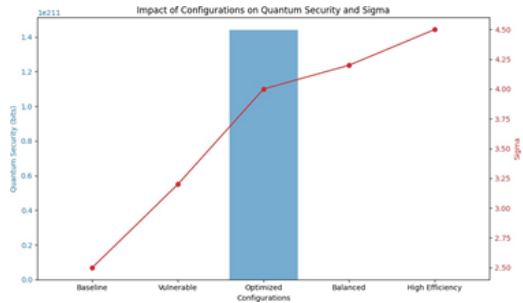

**Figure 3.** *Norm distributions and convergence behavior for N = 1024 configurations, showing the impact of σ on sampling efficiency and security metrics.*

2. ***Security-performance balance***
- Larger σ values improve sampling efficiency but increase lattice ball volume Vol(Bα)
- Smaller σ values ensure higher security at the cost of slower convergence

## Comprehensive analysis of configurations

The relationship between quantum security *(Q_security)* and the Gaussian parameter σ across different lattice dimensions N is summarized in Figure 3. The results highlight key trade-offs between security and efficiency:

- ***Optimized configuration*** (N = 1024, σ = 4.0): Achieves the highest quantum security (7.46×10³⁰¹ bits), ideal for high-security environments.
- ***Intermediate configuration*** (N = 768, σ = 3.5): Provides excellent security (7.47 × 10²²⁴ bits) with balanced performance.
- ***Standard configuration*** (N = 512, σ = 3.5): Offers moderate security (4.11×10¹²⁷ bits) suitable for standard applications.
- ***Lower dimensions*** (N = 256, σ = 2.5): Provides insufficient security (3.72 ×10⁶³ bits), suitable only for testing.

Selecting appropriate parameters is essential for balancing quantum security and computational efficiency in NTRUEncrypt implementations. Aligning with the objectives of parameter optimization for quantum resistance and practical deployment outlined in the introduction.

## Conclusion

This study provides a foundational framework for understanding and improving NTRUEncrypt's quantum resistance. By integrating MCMC methods with lattice cryptography [17], our results offer practical guidelines for parameter selection, secure implementations, and efficient key generation. These findings directly support ongoing standardization efforts for post-quantum cryptography [7].

The analysis demonstrates that parameter optimization is essential for balancing quantum security and computational efficiency [14]. For instance, configurations with N = 1024, σ = 4.0 achieve optimal security (7.46 × 10³⁰¹ bits), while σ = 4.5 provides enhanced performance with robust security (4.81×10²⁷⁵ bits) [11]. Additionally, the Gaussian parameter σ significantly influences the lattice ball volume Vol(Bα), highlighting its exponential impact on security metrics [4].

## Key contributions

Our work achieves several theoretical and practical advancements

***Theoretical insights***
- Rigorous quantum security bounds based on lattice reduction estimates [9]:

$$Q_{security} = \Omega\left(\sqrt{\frac{2^N}{Vol(B_\alpha)}}\right)$$

- Optimal MCMC sampling complexity with proven convergence [16]:

$$T_{total} = O(N^3 \log N \log(1/\epsilon))$$

***Practical recommendations***

Our analysis shows that while N = 1024 configurations provide sub-stantial quantum resistance; lower dimensions may be vulnerable to advanced lattice attacks [21].

## Future directions

Future research should focus on:
- Developing hardware-optimized implementations for resource-constrained environments [8]
- Extending this framework to other lattice-based cryptosystems
- Studying the impact of quantum computing advances on security parameters
- Exploring automated parameter optimization techniques

These results establish NTRUEncrypt as a promising candidate for post-quantum standardization [7], providing both theoretical security guarantees and practical implementation guidelines.